\documentclass[preprint2]{aastex} 
 
\newcommand{\cms}{\hbox{\,${\rm cm^{-2}}$}} 
 
\newcommand{\cmc}{\hbox{\,${\rm cm^{-3}}$}}

\newcommand{\zp}{\hbox{$z_Q^{\prime}$}}
\newcommand{\rp}{\hbox{$r_Q^{\prime}$}}
\newcommand{\zP}{\hbox{$z_P$}}
\newcommand{\rP}{\hbox{$r_P$}}
 
\newcommand{\HII}{\hbox{H\,{\sc ii}}} 
\newcommand{\HI}{\hbox{H\,{\sc i}}} 
 
\newcommand{\OVILB}{\hbox{Ly$\beta$+O\,{\sc vi}}} 

\newcommand{\Lya}{\hbox{Ly$\alpha$}} 
\newcommand{\Lyb}{\hbox{Ly$\beta$}} 
\newcommand{\Lyc}{\hbox{Ly$\gamma$}} 
\newcommand{\afuv}{\hbox{$\alpha_{FUV}$}} 
\newcommand{\nhz}{\hbox{$n^0_{H^0}$}} 
\newcommand{\nh}{\hbox{$n_{H^0}$}} 
\newcommand{\tauv}{\hbox{$\tau_{1160}$}} 
\newcommand{\taue}{\hbox{$\tau_{1160}^{obs.}$}} 
\newcommand{\lao}{\hbox{$\lambda_{obs.}$}} 
\newcommand{\lar}{\hbox{$\lambda_{rest}$}}

\shorttitle{Intergalactic absorption and quasar SEDs} 
\shortauthors{Binette et al.} 
 
\begin{document} 
 
 
\title{Technique for  detecting warm-hot intergalactic gas 
\\in quasar UV spectra
%\altaffilmark{1}
}
 
 
%\author{Luc Binette\altaffilmark{1}, Mario Rodr\'\i guez-Mart\'\i nez\altaffilmark{1}, Sinhue Haro-Corzo\altaffilmark{1} and Isidro Ballinas\altaffilmark{1}}

\author{Luc Binette, Mario Rodr\'\i guez-Mart\'\i nez, Sinhue
Haro-Corzo and Isidro Ballinas}
\affil{Instituto de Astronom\'\i a, 
Universidad Nacional Aut\'onoma de 
M\'exico, Apartado Postal 70-264, 04510 M\'exico,  
DF, Mexico; binette@astroscu.unam.mx.} 
%, and Wei Zheng\altaffilmark{2}} 
%\affil{} 
%\email{aastex-help@aas.orAg} 
 
%XS\author{} 
%\affil{} 
%\email{zheng@pha.phu.edu} 
 
%\and 
 
%\author{R. J. Hanisch\altaffilmark{5}} 
%\affil{Space Telescope Science Institute, Baltimore, MD 21218} 
 
%\altaffiltext{0}{Based on  
%observations with the NASA/ESA Hubble Space Telescope,  
%obtained at the Space Telescope Science Institute, 
%which is operated by the Association of Universities for Research in 
%Astronomy, Inc., under NASA contract NAS5-26555} 
%\altaffiltext{1}{Instituto de Astronom\'\i a, 
%Universidad Nacional Aut\'onoma de 
%M\'exico, Apartado Postal 70-264, 04510 M\'exico,  
%DF, Mexico; binette@astroscu.unam.mx.} 
%\altaffiltext{2}{Center for Atrophysical Sciences, John Hopkins University, 
%Baltimore, MD 21218-2695; zheng@pha.phu.edu.} 
%\altaffiltext{3}{present address: Center for Astrophysics, 
%    60 Garden Street, Cambridge, MA 02138} 
%\altaffiltext{4}{Visiting Programmer, Space Telescope Science Institute} 
%\altaffiltext{5}{Patron, Alonso's Bar and Grill} 

 
\begin{abstract} 
The ionizing spectral energy distribution of quasars exhibits a
steepening of the distribution shortward of $\sim 1200\,$\AA. The
change of the power-law index from approximately $-1$ (near-UV) to
$-2$ (far-UV) has so far been interpreted as being intrinsic to
quasars. We consider the possibility that the steepening may result
from a tenuous absorption component that is anticorrelated with large
mass overdensities. UV sensitive satellites, whose detectors can
extend down to 1000 \AA, can set a useful limit to such an absorption
component through the search of a  flux increase in
the window 1050--1190\,\AA\ (observer frame) with respect to 
an extrapolation of the continuum  above 1230 \AA. Since the recent FUSE or
HST-STIS data do not show any obvious discontinuity in this region,
this effectively rules out the
possibility that intergalactic \HI\ absorption is very important, and
it is concluded that most if not all of the steepening is intrinsic to
quasars.  A smaller flux discontinuity of order 1\% cannot, however,
be ruled out yet and would still be consistent with the warm-hot
intergalactic component if it amounts to 30\% of the baryonic mass, as
predicted by some models of large scale structure formation, provided
its temperature lies around $10^{5.3}$ K.
\end{abstract}

 
\keywords{galaxies: intergalactic medium --- large-scale structure of
 universe --- galaxies: active --- radiative transfer --- 
ultraviolet: general} 
 
%% \citet creates own parenthesis 
%% \citet[hereafter ZKTGD]{ZKTGD} 
%% \citet[see][]{ZKTGD} 
%% \citep inserts citation WITH year (1988)
%% \citep[Paper I]{djo84}  will refer to paper I 
%% \citep{ZKTGD} o \citet{ZKTGD} 
 
\section{Introduction} \label{sec:intro}

The ionizing spectral energy distribution (hereafter ISED) of nearby
active galactic nuclei cannot be observed directly, due to the large
Galactic absorption beyond the Lyman limit. Owing to the redshift
effect, however, we can get a glimpse of the ISED from the spectra of
distant quasars.
%Most studies on the subject indicate a steepening of the distribution
%at wavelengths shorter than $~ 1000\,$\AA\ (see O'Brien et~al. 1988
%and references therein).
The pioneering work of Zheng et~al. (1997, ZK97), using HST-FOS
archived data, identified a noticeable steepening in quasar SEDs near
1200 \AA. The power-law index $\alpha$ ($F_{\nu}\propto
\nu^{\alpha}$) in quasars steepens from $ -1$ for $\lambda >
1200\,$\AA\ to $\simeq -2$ at shorter wavelengths\footnote{Throughout
the text, \lao\ and \lar\ will denote wavelengths in the quasar
rest-frame or in the observer-frame, respectively [$\lar =
(1+z_Q)^{-1} \lao$].}, down to 350\,\AA\ (\lar). In a more recent
study, Telfer et~al. 2002 (hereafter TZ02) found similar results with a
composite spectrum characterized by a mean near-UV index of $-0.7$
steepening to $\simeq -1.7$ in the far-UV.  The above authors favor the
interpretation that the steepening is intrinsic to quasars and that it
is the signature of a comptonized accretion disk.

On the other hand, certain distributions of intergalactic absorption
gas can also cause  an (apparent) steepening of the SED, starting at
1200\,\AA\ (\lar), as shown in this Paper, that is, located in the same
position as the break encountered by TZ02. As long as ISED studies are
based on detectors that do not extend beyond 1200\,\AA\ (\lao), one
cannot readily distinguish between the contribution to the observed
steepening from an intervening absorption model and that of a purely
intrinsic break in the quasar SED as proposed by ZK97 or TZ02.  FUSE
or HST-STIS spectra, however, extend much farther into the UV and will
be shown here to provide us with a compelling test to discriminate
between the two interpretations.

Large scale structure formation models predict that a substantial
fraction of baryons should reside in a warm-hot phase at current
epoch, possibly up to 30--40\% of $\Omega_{bar.}$ \citep{davea}.
Depending on its temperature and distribution with redshift, this
warm-hot gas may contribute to a reduced fraction of the observed steepening
of the ISED. In this Paper, we show that this component may give rise
to a slight flux increase (that is, a discontinuity) in the region
1050--1190\,\AA\ (\lao) with respect to longer wavelengths. The
calibration of the level of this discontinuous excess flux can be used
as a technique to set useful limits for the baryonic mass contribution
from the warm-hot intergalactic component.

The objective (and results) of this Paper is threefold:
\begin{enumerate}

\item Find the simplest and yet physically meaningful absorption gas
distribution which can mimic the observed steepening (\S \ref{sec:steep}),

\item Devise a technique to falsify the previous proposition that
the break is due to absorption, thereby confirming the idea that the
steepening is intrinsic to quasars (\S \ref{sec:false}),

\item Propose a dependable technique to reveal traces of 
a warm-hot baryonic component in the local Universe using UV spectra
(\S \ref{sec:detect}).

\end{enumerate}

\section{Procedure and calculations}\label{sec:cal}

\subsection{Generalized Gunn-Peterson effect} \label{sec:gp}

The Gunn-Peterson (GP) effect \citep{gunn} sets stringent limits for
the presence of neutral diffuse gas at high redshifts. In the simplest
form of the GP test, the absorption gas produces a flux decrement
between \Lya\ and \Lyb. The decrement is measured against a continuum
level, usually taken to be a power-law extrapolated from the region
redward of the \Lya\ emission line. With high resolution spectra, this
technique is sensitive to \HI\ columns otherwise too small to produce
resolvable \Lya\ lines.  \citet{songa} in this manner could set a
limit of the GP opacity $\tau_{GP} < 0.1$ at $z=4.7$. Another
manifestation of the GP effect is illustrated by the staircase
transmission curve of \citet{moeller}, which is the result of
unresolved \Lya\ forest lines in intermediate resolution spectra. This
transmission curve is also characterized by a broad trough (partly due
to photoelectric absorption), the Lyman valley, whose minimum occurs
at a rest-frame wavelength $\sim 650$\,\AA. Before averaging their
quasar spectra, ZK97 and TZ02 statistically corrected each quasar
spectrum for the presence of such a Lyman valley by calculating the
appropriate transmission curve, using the scheme developed by
\citet{moeller}. Finally, the GP effect can be generalized to include
the broadened wings of faint \Lya\ forest lines as a result of the
Hubble flow in underdense regions.

Our contention is that these manifestations of the GP effect are
sensitive to certain classes of \HI\ distributions as a function of
redshift. Other \HI\ distributions might instead be associated to
large scale voids. In that case, the amount of absorption gas would be
decreasing towards the background quasar, against which we are trying
to detect it. Such opacity behavior might be harder to pinpoint, since
it would not necessarily produce a recognizable absorption break at the
rest-frame \Lya. Another factor that could hinder the detection of
intergalactic absorption is that the absorption component evolves in
the opposite direction as the \Lya\ clouds, becoming in other words
progressively more abundant towards lower redshifts.  For instance,
hierarchical structure formation is expected to generate a hot gas
phase, whose mass is predicted to increase with time
\citep[e.g.][]{phillips}, that is, with a temporal evolution  which
is opposite to that of the \Lya\ forest. Any residual absorption it
may cause would occur at much shorter wavelengths, in a wavelength
domain too far from the \Lya\ emission line (\lar) to rely on an
extrapolation of the continuum observed redward of this line.

With these considerations in mind, we have explored various \HI\
distributions, whose absorption behavior is not as evident as that
expected from the traditional GP effect.  
The hypothesis being tested is this Paper is whether all or just a small
fraction of the far-UV steepening in $F_{\nu}$ is caused by
absorption. We will demonstrate in \S\,\ref{sec:res} how this
hypothesis is observationally falsifiable.
Even if most of the SED break turns out to be  intrinsic to quasars,
the technique can still be used to set a lower limit on the
temperature of the warm-hot intergalactic medium (WHIM).

\subsection{Derivation of the transmission curve} \label{sec:tra}

The technique used to generate a simulated  composite spectrum to
compare with the observed spectrum of ZK97 is described in
\citet{binb}.  Briefly, we assume a typical spectrograph sensitivity
window for FOS of 1300--3000\,\AA.  We then simply redshift this
window in locked steps in order to simulate quasars of different
redshifts in the range $0.33 \le z_Q \le 3.6$ (the same range as
ZK97). Before averaging, each redshifted quasar SED is multiplied by
the redshift integrated transmission curve $T_{\lambda}$ (defined
below). We assume the concordance $\Lambda$CDM cosmology with
$\Omega_{\Lambda} = 0.7$, $\Omega_{M}=0.3$ and $h = 0.67$ with $h =
H_0/100$.

To be definite, we adopt a uniform gas distribution  
\citep[see][]{bina}. Equivalent calculations assuming a clumpy
medium were also carried out for comparison. We found that the
ill-defined additional parameters required then do not reveal anything
fundamental or even interesting. Furthermore, the two formalisms are
equivalent for the assumed opacity regime in which the diffuse
absorption component lies in the linear part of the curve of growth,
that is for \HI\ columns $N_{H^0} \ll 10^{12}\,$\cms.

The simulated quasar spectra are divided in energy bins, and for each
quasar rest-frame wavelength bin $j$, we calculate the transmitted
intensity $I_{\lambda_j}^{tr} = I_{\lambda_j} T_{\lambda_j} =
I_{\lambda_j} e^{-\tau(\lambda_j)}$ by integrating the opacity along
the line-of-sight to quasar redshift $z_Q$
\begin{displaymath}
\begin{array}{cc}
\tau(\lambda_j) = \sum_{i=0}^{10}{\int_{0}^{z_Q}
\sigma_i(\frac{\lambda_j}{1+z})\; n_{H^0}(z) \frac{dl}{dz} \; dz}  \label{eq:tau}
\end{array}
\end{displaymath}
where $\lambda_j$ is the quasar rest-frame wavelength for bin $j$ and
$n_{H^0}(z)$ the intergalactic neutral hydrogen density, which
consists of one of the three density distributions discussed below
[eqs (1--3)]. The summation is carried out over the following opacity
sources: photoionization ($i=0$) and line absorption from the Lyman
series of hydrogen ($1\le i \le 10$). Although our code could include
up to 40 levels, we found that considering only the 10 lowest proved
to be adequate.  We adopted a fiducial velocity dispersion $b$ of
30~km/s and assumed a Gaussian profile for the $\sigma_i$ of the
lines.

\subsection{Heuristic \HI\ gas distributions} \label{sec:hi}

\subsubsection{Three \HI\ distributions to test} \label{sec:three}

In calculating the transmission function of a quasar at redshift
$z_Q$, we will consider three different distributions of absorption
gas density, \nh, with redshift, which we will then compare with the composite
SED of either ZK97 or TZ02. The first is given by
 
\begin{equation}
\label{eq:nha}
\begin{array}{cc}

 n_{H^0}(z) = n^0_{H^0} \,(1+z)^3 \, {(1+z)^{\gamma}} \, {(z_Q^{\prime})^{\beta}}  

\end{array}
\end{equation}
where $z$ is the absorbing gas redshift, $z_Q$ the quasar redshift,
\zp\ the quasar redshift as seen from the absorbing gas at
$z$ [that is $z^{\prime}_Q = (1+z_Q)/(1+z)\,-1$], and \nhz\ the
neutral gas density at zero redshift. $\gamma$ and $\beta$ are
adjustable parameters.  The $(1+z)^3$ factor represents the
cosmological expansion of the Universe.  The factor
$(z_Q^{\prime})^{\beta}$ with $\beta > 0$ implies that the density
decreases in the neighborhood of individual quasars.

The second distribution which will be considered is
\begin{equation}
\label{eq:nhb}
\begin{array}{cc}

 n_{H^0}(z) = n^0_{H^0} \,(1+z)^3\, {\frac{ {(1+z)^{\gamma}} }{1+({{z_P}/{z_Q^{\prime}}})^{\beta} }}

\end{array}
\end{equation}
where \zP\ is the size of the region (near each quasar) within which
the density decreases as $({z_P}/{z_Q^{\prime}})^{\beta}$ towards the
quasar [note that $\beta > 0$ as in eq.~(\ref{eq:nha})].  It is
possible to substitute the denominator above by the expression
${1+({{r_P}/{r_Q^{\prime}}})^{\beta}}$. For each value of \zP\
discussed below, we we have searched for the value of \rP\ which would
give us an equivalent fit of the composite ISED. The parameter \rP\ represents the
zone of influence (or cavity radius) of the quasars in spatial units.
Although the density decreases smoothly towards each quasar, in this
text we will refer to this zone, where $\rp < \rP$ near each quasar,
as a `cavity'.

The third distribution is given by

\begin{equation}
\label{eq:nhc}
\begin{array}{cc}

 n_{H^0}(z) = n^0_{H^0} \,(1+z)^3 \, {\frac{ {\rm exp}[-(z/1.6)^{1.4}] }{1+({{z_P}/{z_Q^{\prime}}})^{\beta} }}

\end{array}
\end{equation}
where the exponential function in the numerator is a parametric fit to
the increase of the warm-hot intergalactic medium (WHIM) towards the
current epoch as calculated by \citet{davea} (their Model D2). We will
again discuss which value of \rP\ produces a fit equivalent to that
provided by our favored \zP\ values.

\subsubsection{Three conditions to satisfy} \label{sec:cond}

The above distributions lie in a sequence of increasing plausibility,
as discussed below. We emphasize that our exploration included many
more functions than the above three, the vast majority of which could
not reproduce a smooth steepening of the spectrum near 1100 \AA\
(\lao) (failing condition {\it a} below). Nevertheless, these three
distributions will suffice for the purpose of unequivocally testing
the absorption hypothesis.

In our search, the following procedure was adopted: we explored a
large set of functions with the following three priorities in mind:
{\it a)} the \nh\ distribution must produce an acceptable fit of the
the composite ISED without a discontinuity near
\Lya, {\it b)} it must be a physically meaningful function, 
and {\it c)} it must be consistent with other information we have
about the Universe such as, for instance, the volume density of quasars.
We were able to satisfy conditions {\it a} and {\it b}, although {\it
c} could not be strictly respected for our favored test case Model C,
as discussed below.

\section{Model results} \label{sec:res}

\subsection{Reproducing the observed steepening} \label{sec:steep}
 
If we adopt the first distribution of \HI\ corresponding to
eq.~(\ref{eq:nha}) and the parameters listed in Table~\ref{tbl_1} for
Model A, that is, a zero redshift density $\nhz = 4.7 \times 10^{-12}
$\,\cmc, $\gamma=-1.5$ and $\beta = 0.8$, we can reproduce the
observed smooth steepening of the composite spectrum of ZK97
remarkably well as shown in Fig.~\ref{fig1} by the thick solid
line. The intrinsic SED considered in Model~A is a single power-law of
index $-1$ (short-long dashed line), which we extend to all
wavelengths. This index provides a good fit at the longer wavelengths
(1300--2300\AA) of the underlying continuum in the ZK97 composite
spectrum.

This first distribution has also the property that the far-UV index,
\afuv, defined in the range 350--900 \AA\ (\lar), increases very little
with quasar redshift \citep{bina}. This agrees with the findings of
TF02 (c.f. their Fig.~12). However, although the simplest, this
distribution is not physically meaningful in the following aspect: the
limitless range of the dependence of eq. (\ref{eq:nha}) on \zp\
implies that the local \HI\ density depends on the quasar redshift. To
clarify this point: two independent quasars, one nearby and the other
at redshift 3, but lying along very nearby line-of-sights on the sky,
would be characterized by very different local \HI\ densities according
to eq. (\ref{eq:nha}).  Such behavior is unphysical, because the \HI\
density in the local Universe should not depend at what distance the
background quasar is located.  We therefore conclude that condition {\it b}
(`physically meaningful', see \S \ref{sec:cond}) is not satisfied and
that the \nh\ distribution of eq. (\ref{eq:nha}) must be rejected. In
the text, we will refer to this undesirable property as the ``nearby
line-of-sight problem''.

A distribution without any dependence on \zp\ would not suffer from
this problem. However,  in {\it all} the distributions
that succeeded in reproducing the smooth roll-over near 1200
\AA,  some kind of smooth gas cavity near each quasar
had to be taken in, otherwise, a deep and sharp absorption
discontinuity\footnote{In \citet{binb}, such a discontinuity at 1216
\AA\ (\lar) was smoothed out by invoking photoionization by the
background quasar (in analogy with the ``proximity effect''). However,
the metagalactic background radiation (relative to a single quasar) is
much too strong for this proposition to be sustained. \citet{eastman}
avoided the discontinuity by invoking absorption cloudlets being
progressively accelerated up to $0.8 c$ within a region $< 10$\,kpc
near the background quasar. The main problem was that the cloudlets
themselves emitted more UV than the background quasar.} unavoidably
appears at 1216 \AA\ (\lar) in the synthesized composite SED, a
feature that is obviously not observed.  The thin long-dashed line in
Fig.~\ref{fig1}, for instance, illustrates the magnitude of this sharp
discontinuity present near \Lya\ when no cavity is considered (here
with $\beta$ set to zero in eq. (\ref{eq:nha})).

The solution to this problem is to have the dependence on \zp\ limited
to a reduced zone of influence near each quasar. This is achieved with
the expression in the denominator in eqs~(\ref{eq:nhb}) and
(\ref{eq:nhc}), where \zP\ limits the distance scale within which the
density is decreasing (towards the quasar) as a power-law of index
$\beta$.  We therefore have two competing effects present in our
subsequent distributions [of eq. (\ref{eq:nhb}) or
eq. (\ref{eq:nhc})]: one factor describes the \HI\ density as a
function of redshift (seen from us) $z$, which has no other QSO close
to the selected line-of-sight, and the other describes the \HI\
density within the neighborhood of QSOs (within a zone of influence
$\simeq \rP$). Two different line-of-sights lying at a projected
distance on the sky greater than \rP\ will be described by a unique
and self-consistent distribution of \HI\ with redshift. As a consistency
check, the volume occupied by all the QSO cavities must be smaller
than the total volume sampled.

A possible interpretation of the decreasing density towards individual
quasars is that our putative \HI\ distribution is associated to the
largest scale voids. The environment of quasars might also be hotter
and transparent as a result of (protocluster) stellar winds from the
associated large scale and most massive structures. Our aim will be to
make \zP\ as small as possible, but without producing any
discontinuity or dip near 1200 \AA. The short-dashed line in
Fig.~\ref{fig1} shows that the distribution corresponding to
eq.~(\ref{eq:nhb}) (Model B) results in as good a fit to the ZK97
spectrum as Model A, when using $\gamma = -3.0$, $\zP = 0.7$ and $\beta =
1.5$ (see Table~\ref{tbl_1}).

Nevertheless, this Model B is not satisfactory.  In effect, a cavity
with $\zP = 0.7$ requires a cavity of size $\rP
\sim 2$ Gpc, which is more than half the path-length to the
highest-$z$ quasar in the sample (i.e. 3.75 Gpc for $z_Q=3.6$). Hence
most line-of-sights would cross many overlapping cavities (hence
condition {\it b} is not satisfied) and such a situation not properly
described by eq.  (\ref{eq:nhb}). Condition {\it c} (`consistency with
other reliable information', see \S \ref{sec:cond}) is not respected
either. First, the cumulative volume of quasar influence on \HI\
exceeeds 15\% of the total volume sampled. Second, the exponent of
$(1+z)$, $\gamma = -3.0$, is much too steep if we were to compare our
Model B to model-predictions of the evolution of the WHIM. For
instance, most models in \citet{davea} are characterized by
redshift-averaged values of $\gamma$ in the range $-1.1$ to $-1.5$.

In order to ensure a physically more meaningful \HI\ distribution, we
replaced the ($\gamma$) power-law dependence  on redshift by an
exponential fit to the WHIM Model D2 of \citet{davea}, the description of
which lies in the numerator of eq.~(\ref{eq:nhc}).  Replacing the
previous best fit distribution with $\gamma =-3$ by a distribution
that  matches the behavior of the WHIM with redshift better [by using
either eq.~(\ref{eq:nhc}) or $\gamma \simeq -1.2$ and eq.~(\ref{eq:nhb})]
comes at a price, however. In effect, such distributions always result
in a smooth rise in transmission towards the short wavelength
extremity of the SED, beyond 500 \AA. Interestingly, the more recent
composite spectrum from TZ02 presents such a rise in the far UV as
displayed in Fig.~\ref{fig2}. TZ02 have questioned the reality of this
rise. The latter might be caused by a tendency of high redshift
quasars to have a harder ISED. Nonetheless, because the TZ02 work
contains more objects than the ZK97, and because this feature is seen
in both radio-loud as well as radio-quiet quasars, we will take their
results at face values and compare their radio-quiet composite SED
with models which used eq.~(\ref{eq:nhc}). The thick solid line in
Fig.~\ref{fig2} represents our Model~C, which is characterized by a
significantly smaller $\zP = 0.3$. Although the fit is imperfect, it
matches the overall trends present in the TZ02 composite spectrum. The
adopted intrinsic SED used here is somewhat harder, corresponding to
an index of $-0.72$ (which is the value inferred by  TZ02).

For Model C, with $\zP = 0.3$ we find that $\rP = 800$ Mpc. A single
cavity  occupies therefore 1\% of the total volume sampled. Using the
known luminosity function of bright QSOs [i.e. a quasar density of
order $\xi = 3 \times 10^{-7} \, {\rm Mpc}^{-3}$ \citep{Boyle}], the
volume\footnote{That is, one should realize the condition $V_Q =
\frac{4}{3} \pi {\rP}^{3} \la \xi^{-1}$\,.} 
corresponding to this value of \rP\ is too large by a factor $\sim 650$
and condition {\it c} is therefore not fulfilled on the ground of the
known density of QSOs. With this caveat in mind, we proceed to study
the behavior of the absorption of our putative intergalactic
component at wavelengths shorter than provided by HST-FOS. This will
allow us to calibrate a new method to detect the WHIM and, indirectly,
to determine to what extent condition {\it b} is satisfied in Model C.

\subsection{Prediction of a possible discontinuity in the far UV} \label{sec:false}

\subsubsection{Analysis of the transmission curves}  \label{sec:ana}

The derivation of the composite SED of Model C required using
eq. (\ref{eq:nhc}) and calculating the transmission function at evenly
spaced redshift values.  We now turn to the particularities of these
transmission curves.  Two such transmission curves are presented in
Fig.~\ref{fig3} for illustrative purposes for quasars of redshifts
2.76 and 0.33. Clearly, the main opacity source is the \Lya\ line. The
contribution of photoelectric absorption to the total opacity is
relatively small as illustrated in the Figure while the contribution
of higher series lines can be appreciated by the reduced size of the
jumps that are visible in the transmission curves. The main result to
emphasize is that a discontinuity appears to the blue of \Lya\ (\lao)
(that is beyond the UV limit of the HST-FOS detector) in all the
calculated transmission curves, not only when assuming the parameters
of Model C, but in all the \HI\ distributions that satisfied condition
{\it a} (`good fit', see \S \ref{sec:cond}).

By comparing the transmission curves in Fig.~\ref{fig3} with those
resulting from \Lya\ forest absorbers, a clear difference emerges. The
transmission functions derived by \citep{moeller} (or ZK97) are
characterized by a stair-case behavior, in which the opacity from each
physical process (either a Lyman series line or photoelectric
absorption) {\it decreases} towards shorter wavelengths, due to the
strong evolutionary nature of the \Lya\ forest, in which the density
of absorbers increases steeply with redshift. In the case of the
transmission resulting from equations (1)--(3), the opacity (along a
given step of the staircase transmission curve) tends to either remain
high or increase towards shorter wavelengths (but, of course, only up
to the threshold wavelength of the absorption process involved). The
obvious reason for this marked difference is that the \HI\
distributions used here are characterized by a density that increases
towards smaller redshifts $z$.

An inspection of Fig.~\ref{fig3} suggests that, rather than looking
for a sharp absorption edge near \Lya\ in the quasar rest-frame, the
most obvious demarcation produced by \nhz\ distributions (which
satisfy condition {\it a} of a `good fit') is at the shorter
wavelength end, in the 1190--1070 \AA\ observer-frame region where
\Lyb\ rather than \Lya\ dominates the absorption.  In order to detect
such a discontinuity near 1216 \AA\ (\lao), one requires detectors
with a sensitivity, which extends to wavelengths shorter than the
earlier HST-FOS window, such as provided by FUSE or HST-STIS.

\subsubsection{Operative measurement of the jump: \tauv }  \label{sec:def}

In order to test our absorption hypothesis against individual quasar
spectra observed by STIS or FUSE, we must first calibrate the depth of
the expected discontinuity (on the blue side of local \Lya) as a
function of \nhz\ for our different models.  Since the intrinsic far-UV
spectral index, \afuv, is not precisely known and, as suggested by the
substantial scatter found by TZ02, may even vary from quasar to
quasar, it makes sense to measure the depth of the jump in a way that
does not depend on an a priori knowledge of the intrinsic ISED. The
technique proposed here is to evaluate the discontinuity's depth at
1160\,\AA\ (\lao) by comparing the flux there to the extrapolated
value from a power-law fit redward of \Lya\ (\lao), within the narrow
window 1260--1360\,\AA.  This window is meant to exclude the Galactic
\Lya\ absorption trough as well as the geocoronal \Lya.  We define the
quantity $\tauv = {\rm log_e} \, [F^{1160}_{obs.}/F_{extr.}^{1160}]$,
which can be shown to be insensitive to the power-law index of the
intrinsic ISED assumed. In Fig.~\ref{fig3}, we illustrate the
procedure by taking as example the transmitted SED of a $z_Q=1.0$
quasar and Model C (the calculation assumes an ideal detector without
wavelength coverage limitations).  The two squares illustrate the
position (albeit here in the quasar rest-frame), at which it is
proposed to define \tauv. For each model discussed (A--D), we list in
Table~\ref{tbl_1} the value of \tauv\ evaluated at $z_Q =1$, while in
Fig.~\ref{fig4} we plot the behavior of \tauv\ as a function of $z_Q$.

\subsubsection{Condition b and the  behavior of \tauv }  \label{sec:phy}

As discussed in \S \ref{sec:steep} concerning the ``nearby
line-of-sight problem'', it is unphysical to have the local density
vary when the background quasar, for instance, lies at redshift 2
rather than 3. We can easily find out when a distribution suffers from
this problem by checking whether or not \tauv\ is constant at moderate
and high values of quasar redshift.  It is apparent in Fig.~\ref{fig4}
that Models A and B suffer severely from the above-mentioned problem.
In the case of Model C, however, \tauv\ is flat, at least beyond $z_Q
\ga 0.6$. However,
within a radius around us given by the large value of $\rP \sim 800$
Mpc, we may reasonably expect to lie within the zone of influence of a
(dominant) quasar.  In this case, rather than the smoothly increasing
function depicted in Fig.~\ref{fig4}, a single value of
\taue\ may   instead apply (\tauv\ reflects  the local \HI\
density).  Its value would depend on our distance from this dominant
quasar.  Furthermore, \tauv\ would not necessarily be isotropic. In
conclusion, the smooth initial rise of \tauv\ in Model C is at best an
idealization. Given the likelihood of being positioned inside the
cavity of a single quasar, the predicted \tauv\ constitutes an upper
limit (a more probable value may lie a factor of a few lower).

\subsubsection{Negative results using two EUV quasar  spectra}  \label{sec:neg}

We now use Model~C as a test case, since it is the most physical at
hand under the assumption that the 500 \AA\ rise in the composite ISED
of TZ02 is real and that the break is entirely due to intergalactic
absorption.  For any quasar $z_Q > 0.5$, one finds that \tauv\ is
quite significant. Typically, $\tauv \ga 0.25$ as seen in Fig.\ref{fig4}. If
that much absorption were present, it would therefore be a striking
feature in the far-UV spectra [provided the quasar could be observed
blueward of 1200 \AA\ (\lao)]. It turns out, however, that the
observations of the quasar HS1543+5921 by
\citep[][$z_Q=0.807$]{bowen} with STIS and HE2347--4342 by
\citep[][$z_Q=2.885$]{kriss}  with FUSE, show no
indication at all of a discontinuity at the expected level of $\tauv
=0.27$ and 0.28, respectively, in the 1160 \AA\ region, as compared to
the extrapolation from the 1270 \AA\ region. We consider that a
discontinuity as small as $\tauv = 0.05$, if present in the two
spectra, would have been apparent.

We conclude that intergalactic absorption cannot be the main cause of
the observed steepening in the composite SED. Model C fails not only
because of the observed absence of a discontinuity in the far-UV but
also because it is inconsistent with the known density of quasars
(condition {\it c}). No other \HI\ distribution could be found that
would solve alll three conditions.  Our study therefore confirms that
the change of slope near 1100 \AA\ must be in origin intrinsic to
quasars, as proposed by ZK97 and TZ02. This general conclusion is not
dependent on the particular distribution adopted, since in all the
distributions we have explored (which satisfied condition {\it a}),
\tauv\ was always $\ge 0.2$ when $z_Q > 0.5$.

\subsection{Detecting the WHIM in absorption} \label{sec:detect}

Having  recognized that most of the break is intrinsic to quasars,
we now turn to the  problem of determining how much flux
discontinuity near 1160 \AA\ (i.e. \tauv) can be expected if we assume
the \HII\ density predicted by WHIM models and a fiducial temperature
of $10^{5.3}$ \,K.  For this purpose, we adopt an intrinsic SED, which
already incorporates the intrinsic far-UV steepening. It consists of a
broken power-law, which has the same index of $-0.72$ in the near-UV
as in Fig. \ref{fig2}, but that sharply turns over at 1200 \AA\ (\lar)
into a steeper index of $-1.57$ in the far-UV, as represented by the
short-long dashed line in Fig. \ref{fig5}.  These two indices
correspond to the values characterizing the radio-quiet quasar
composite of TZ02 while the turn-over wavelength is within the range
of values encountered by TZ02 ($\approx $ 1200--1300 \AA).

\subsubsection{Model D using a broken power-law SED}

Adopting eq.~(\ref{eq:nhc}) and the above broken power-law, we first
determine which values of \nhz\ and \zP\ best reproduce the observed
composite SED. The result is Model D, plotted as the solid line in
Fig.~\ref{fig5}.  The same $\beta=1.5$ is used as for Model C and \zP\
could now be set to the small value of 0.08 without producing any significant
discontinuity\footnote{The value of \rP\ which can provide an
equivalent fit is 200 Mpc. Interestingly, this value is only a factor
of two too large with respect to the density of quasars at high
redshifts. This can be resolved if we use a  function \rP\ of
redshift such as $300 \,(1+z)^{-1}$\, Mpc. Although \rP\ is then larger in the
local Universe, at redshifts $\sim 2$ we  get $\rP\ \simeq 100$
\, Mpc, which is the epoch when  the density of quasar is known to
peak. Condition {\it c} is marginally satisfied in Model D.}  on the
blue side of the quasar \Lya\ emission line.  Model D succeeds rather
well in fitting the TZ02 composite. It does not constitute, however, a
unique solution because of the uncertainties regarding the intrinsic
SED. If for instance, we shifted the break from 1200 to 1100 \AA, the
assumed ISED would lie higher above the composite of TZ02 and a higher
density WHIM would therefore be required for the model to overlap the
data.

With a predicted value of \tauv\ as small as $0.05$ (see Table
\ref{tbl_1}), Model D is characterized by a much
smaller discontinuity than Models A--C, simply because the assumed ISED
is much closer to the observed composite of TZ02. Furthermore, \tauv\
does not depend  on redshift anymore, as shown in Fig. \ref{fig4}, which
indicates that the cavity near the individual quasars is not affecting
our determination of \nhz.  Inspection of the published spectra of
HS1543+5921 by \citep[][]{bowen} and HE2347--4342 by \citep[][]{kriss}
suggests, however, that \taue\ is likely to be smaller than 0.05.
A thorough analysis   aiming at  setting stringent upper limits on
\taue\ would  clarify the matter.

\subsubsection{Deriving \tauv\ from the expected WHIM} \label{sec:wh}

Let us now turn to the inverse problem of determining the height of
the jump \tauv\ expected from a WHIM component that would contribute
30\% of the total baryonic mass in the local Universe
\citep[as in Model D2 of ][]{davea}.  One advantage of the operative
definition of \tauv\ proposed in \S \ref{sec:def} is that the choice
of a given intrinsic SED is not critical to the purpose of determining
\nhz.  This is verified from Table \ref{tbl_1} where the quotient
$\nhz/\tauv$ evaluated at $z_Q=1$ in Models C and D are $1.89 \times
10^{-11} $ and $1.80 \times 10^{-11} \,$ \cmc, respectively. They
differ by only 5\%, despite the use of very different intrinsic
SEDs. Because \nhz\ scales linearly with \tauv, all the information
needed to derive \tauv\ can be extracted from Model D independently of
whether or not the density inferred turns out very different than in
Table \ref{tbl_1} or whether the ISED assumed is somewhat incorrect.

For any arbitrarily small {\it observed} discontinuity
\taue, the inferred \HI\ density is  given by 
$\nhz = 1.8 \times 10^{-11} \, \taue$\,\cms\ which does not depend
on $z_Q$ because we have $\zP \ll 0.4$.  In terms of the critical
density $3H_0^2/8\pi G m_H$, this density becomes
\begin{equation}
\begin{array}{cc}
\Omega_{WHIM} =  {{1.6 \times 10^{-6}}\over{x \, h^2}} \; \taue \label{eq:om}
\end{array}
\end{equation}
where $x$ is the neutral fraction of the putative WHIM absorption gas
component. We can express $\Omega_{WHIM}$ as a fraction $F_b$ of the
(current epoch) total baryonic mass $\Omega_{bar.} = 0.021 h^{-2}$ 
\begin{equation}
\begin{array}{cc}
F_b  = \Omega_{WHIM}/\Omega_{bar.} = {{7.6 \times 10^{-5} }\over{x}} \; \taue  \label{eq:fb}
\end{array}
\end{equation}
With $F_b=0.3$, we obtain that $\taue \simeq x/2.5 \times
10^{-4}$.  Since the WHIM is likely to be highly ionized, we are lead
to expect very small values of \taue.  For example, for a
collisionally ionized WHIM at a temperature $10^{5.3}$\,K, we derive
$x \sim 3.1 \times 10^{-6}$ and therefore $\taue = 0.012$. Clearly, a
high S/N spectrum is required in order to be able to detect a 1\%
excess flux at 1160\,\AA. It remains nevertheless a worthwhile
exercise to carry out, since any limit that can be put on \taue\ would
directly translate into a lower limit on the neutral fraction (hence
on the temperature) of the WHIM component. Note that, with that small
\taue, we have the liberty to set \zP\ to zero in eq. (\ref{eq:nhc})
and condition {\it c} (\S \ref{sec:cond}) is therefore implicitly
satisfied. By the same token, we do not have to worry about whether we
possibly lie in the zone of influence of a single quasar anymore. In
order to optimize the detection sensitivity, the strategy to follow
would be to add together as many EUV quasar spectra as possible
(ideally with redshift in the range $0.5 \la z_Q \la 0.8$) and then
extract \taue\ from the summed stack.

\section{Conclusions} \label{sec:conc}

In all the distributions of $\nh(z)$ used for reproducing the
steepening observed in the energy distribution of quasars, the
dominant atomic process contributing to the opacity is \Lya\ 
scattering. An important implication of these calculations is that the
transmission curve shows a significant discontinuity to the blue of
\Lya\ (\lao) where the transmission rises sharply. The size of the
discontinuity, \tauv, for any distribution of intergalactic gas that
can successfully reproduce the SED steepening reported by TZ02
(assuming a single power-law SED), is predicted to be typically $>
0.25$ in spectra of quasar redshift $z_Q > 0.5$. Since the two quasars
HS1543+5921 and HE2347--4342, which were observed by HST-STIS and FUSE,
respectively, do not show any sign of a discontinuity anywhere near this level,
we consider  the case proven that most if not all of the 1200 \AA\
continuum steepening is intrinsic to quasars rather than due to \HI\
absorption. On the other hand, a small flux increase by as little as
1\% percent is still possible and  would be consistent with a WHIM
temperature $\sim 10^{5.3}$\,K contributing up to 30\% of the
current baryonic matter.

\begin{figure} 
\plotone{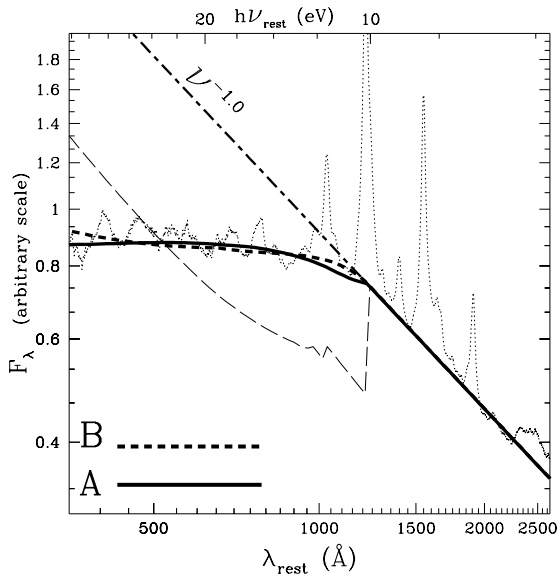} 
\caption{The dotted-line in this  $F_{\lambda}$ vs \lar\ plot
represents the composite quasar spectrum of ZK97.  The straight
short-long dashed line represents a power-law fit, in the longer
wavelength region $\lambda > 1200$\,\AA\ (\lar), of the continuum
 underlying the emission lines. The thick solid line overlaying
the ZK97 data is our simulated composite Model~A, which assumes a
single power-law with $F_{\nu} \propto \nu^{-1}$ to describe the
intrinsic quasar SED and a tenuous
\HI\ absorption  screen  defined by  eq. (\ref{eq:nha}) with
 the parameters listed in Table~\ref{tbl_1}. The short-dash line
represents Model~B, in which the \HI\ distribution corresponds to
eq. (\ref{eq:nhb}) with $\zP = 0.7$. The thin long-dashed line
illustrates how deep the discontinuity at $\lao =1216$
\AA\ is in models   without \zp\ dependence  (i.e. without cavity, the curve
shown was obtained by setting $\beta = 0$  in Model A).
\label{fig1}}
\end{figure}

\begin{figure} 
\plotone{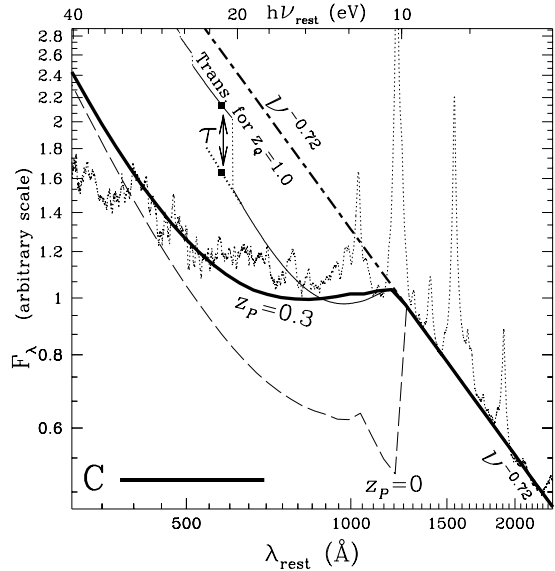} 
\caption{The dotted-line in this  $F_{\lambda}$ vs \lar\ 
plot represents the composite spectrum of radio-quiet quasars by TZ02.
The straight short-long dashed line represents a power-law fit (of
index $-0.72$), in the longer wavelength region $\lambda > 1200$\,\AA\
(\lar) of the continuum  underlying the emission lines. The
thick solid line overlaying the TZ02 data is our composite Model~C,
assuming a single power-law, $F_{\nu} \propto \nu^{-0.72}$, for the
intrinsic quasar SED, and a tenuous \HI\ absorption screen described
by eq. (\ref{eq:nhc}) with the parameters listed in Table~\ref{tbl_1}
($\zP = 0.3$). The staircase thin continuous line represents the
transmitted spectrum for an {\it individual} quasar of redshift $z_Q = 1.0$,
assuming eq.~(\ref{eq:nhc}) [Model C] and an ideal detector with
sensitivity at all wavelengths. The natural logarithm of the flux jump
between the two filled squares is \tauv\ (see definition in
\S\,\ref{sec:def}). The thin long-dashed line illustrates the 
discontinuity at $\lar =1216$ \AA\  if we set
\zP\ to zero in Model C (i.e. without cavity).
\label{fig2}}
\end{figure}

\begin{figure} 
\plotone{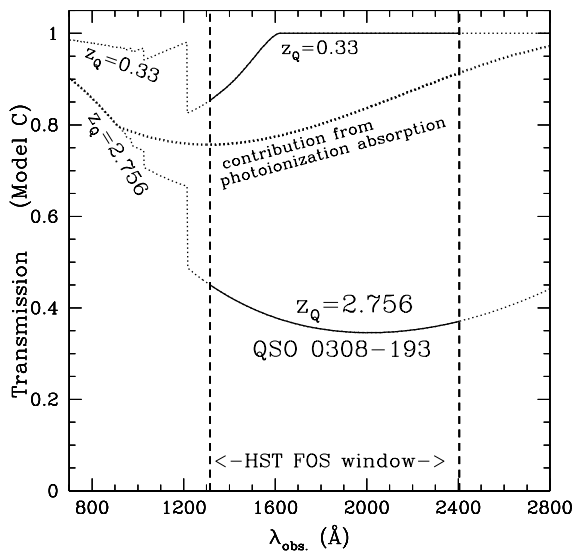} 
\caption{The lower solid line is the calculated transmission curve for
the quasar 0308$-$193 at $z_Q= 2.756$ as a function of the {\it
observer-frame} wavelength for Model~C [see eq. (\ref{eq:nhc}) and Table
\ref{tbl_1}].  The dotted lines shows the continuation of the
transmission curve outside the HST-FOS spectral window used on that
occasion (1315--2400 \AA). Starting from the right, the first, second
and third ramps, seen in the transmission curve, are dominated by the
opacity of \Lya, \Lyb\ and \Lyc\ lines, respectively. The line labeled
``contribution...'' illustrates the small contribution of
photoionization to the total opacity.  The upper solid line is the
transmission curve for a lower redshift quasar at $z_Q= 0.33$.
\label{fig3} }
\end{figure}

\begin{figure} 
\plotone{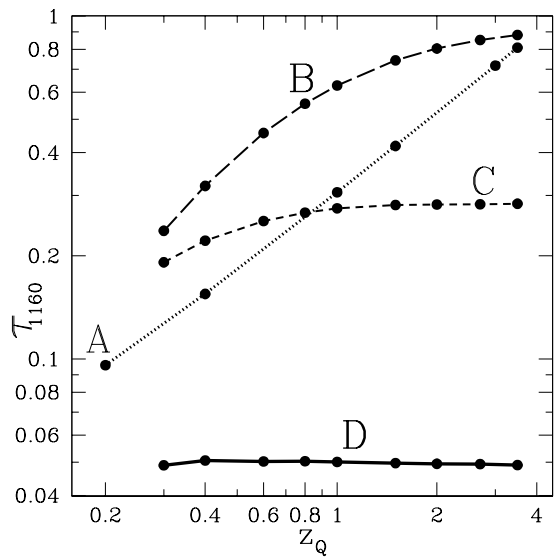} 
\caption{ Behavior of the depth of the flux discontinuity, \tauv, evaluated  
at 1160 \AA\ (\lao) (c.f. \S\,\ref{sec:def}), as a function of quasar
redshift $z_Q$ for the Models A, B, C and D.
\label{fig4} }
\end{figure}

\begin{figure} 
\plotone{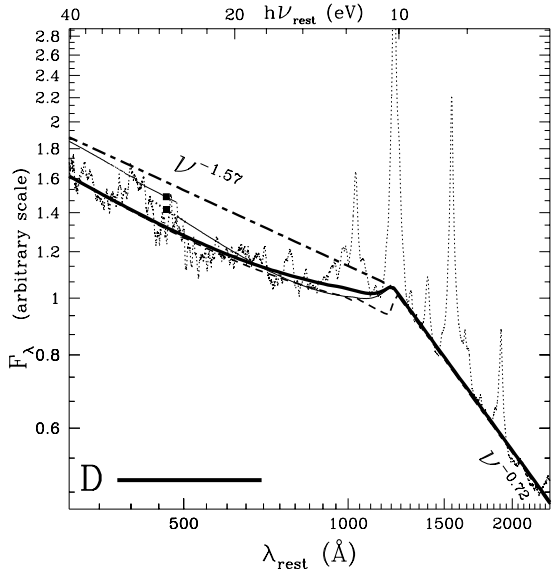} 
\caption{The dotted-line in this  $F_{\lambda}$ vs \lar\ 
plot represents the composite spectrum of radio-quiet quasars by TZ02.
The straight short-long dashed line represents the intrinsic
broken power-law SED assumed in  Model~D. It has a break at 1200 \AA\ at which
point the power-law index changes from $-0.72$ to $-1.57$. The thick
solid line overlaying the TZ02 data is Model~D using the parameters
listed in Table~\ref{tbl_1}. As in Fig. \ref{fig2}, the thin
continuous line represents the transmitted spectrum of an {\it
individual} quasar with $z_Q = 1.0$ (using  Model~D parameters).
The thin short-dashed line illustrates the discontinuity at $\lar
=1216$ \AA\ if we set \zP\ to zero in Model D (i.e. without cavity).
\label{fig5}}
\end{figure}

\acknowledgments 
This work was supported by the Mexican Science Funding Agency CONACyT
under grant 32139-E. We are indebted to Wei Zheng and Randal Telfer
for sharing their published composite spectrum which are used in
Fig.~\ref{fig1} and \ref{fig2}, respectively. We are grateful to
V. Avila-Reese for constructive discussions about the manuscript. 
 
%%% We have no observations of \ion{Ca}{2}. 
%\url{http://www.aas.org/publications/aastex} or the 
%\anchor{ftp://www.aas.org/pubs/}{AAS ftp site}. 
%For technical support, please write to 
%\email{aastex-help@aas.org}. 

%\clearpage  

\clearpage

\begin{deluxetable}{ccllcccc}
\tabletypesize{\scriptsize}
\tablecaption{Model parameters \label{tbl_1}}
\tablewidth{0pt}
\tablehead{
\colhead{Label} & \colhead{Equation}   & \colhead{SED}   &
\colhead{$\nhz(z=0)$} & \colhead{$\gamma$}  &
\colhead{$\beta$}  & \colhead{\zP} & $\tauv$ \\
 & \colhead{(\#)}   & \colhead{$\lambda < 1200$ \AA}   &
\colhead{(\cmc)} &   &   &   &  ($z_Q=1$)
}
 \startdata
A & (1) & $\nu^{-1}$ & $ 4.7 \times 10^{-12} $ & $-1.5$ & 0.8 & --- &
0.306 \\
B & (2) & $\nu^{-1}$ & $ 1.5 \times 10^{-11} $ & $-3.0$ & 1.5 & 0.70 &
0.628 \\
C & (3) & $\nu^{-0.72}$ & $ 5.2 \times 10^{-12} $  & -- & 1.5 & 0.30 &
0.275 \\
D & (3) & $\nu^{-1.57}$ & $ 9.0 \times 10^{-13} $  & --  & 1.5 & 0.08 &
0.050 \\
 \enddata
%% Text for table notes should follow after the \enddata but before
%% the \end{deluxetable}. Make sure there is at least one \tablenotemark
%% in the table for each \tablenotetext.
%%\tablenotetext{a}{Where $h = H_0/100$. The calculations were originally done using $h =0.67$.  }
%\tablecomments{Occasionally, authors wish to append a short
%paragraph of explanatory notes that pertain to the entire table, but }
\end{deluxetable}


\clearpage

%\clearpage  

%\clearpage  

%% Tables should be submitted one per page, so put a \clearpage before 

\end{document}